\RequirePackage{ifpdf}
\ifpdf % We are running pdfTeX in pdf mode
\documentclass[pdftex]{sigma}
\else
\documentclass{sigma}
\fi

\numberwithin{equation}{section}

\begin{document}

\allowdisplaybreaks

\renewcommand{\PaperNumber}{096}

\FirstPageHeading

\renewcommand{\thefootnote}{$\star$}

\ShortArticleName{Restricted Flows and the Soliton Equation with
Self-Consistent Sources}

\ArticleName{Restricted Flows and the Soliton Equation \\ with
Self-Consistent Sources\footnote{This paper is a contribution to
the Vadim Kuznetsov Memorial Issue ``Integrable Systems and
Related Topics''. The full collection is available at
\href{http://www.emis.de/journals/SIGMA/kuznetsov.html}{http://www.emis.de/journals/SIGMA/kuznetsov.html}}}

\Author{Runliang LIN~$^\dag$, Haishen YAO~$^\ddag$ and Yunbo ZENG~$^\dag$}

\AuthorNameForHeading{R.L. Lin, H.S. Yao and Y.B. Zeng}

\Address{$^\dag$~Department of Mathematical Sciences,
    Tsinghua University, Beijing 100084, P.R. China}
\EmailD{\href{mailto:rlin@math.tsinghua.edu.cn}{rlin@math.tsinghua.edu.cn},
\href{mailto:yzeng@math.tsinghua.edu.cn}{yzeng@math.tsinghua.edu.cn}}
\URLaddressD{\url{http://faculty.math.tsinghua.edu.cn/faculty/~rlin/}}

\Address{$^\ddag$~Dept. of Math and Computer Science, QCC, The
City University of New York, USA}
\EmailD{\href{mailto:hyao@qcc.cuny.edu}{hyao@qcc.cuny.edu}}

\ArticleDates{Received October 28, 2006, in f\/inal form December
22, 2006; Published online December 30, 2006}

\Abstract{The KdV equation is used as an example to illustrate the
relation between the restricted f\/lows and the soliton equation
with self-consistent sources. Inspired by the results on the
B\"acklund transformation for the restricted f\/lows (by V.B.
Kuznetsov et al.), we constructed two types of Darboux
transformations for the KdV equation with self-consistent sources
(KdVES). These Darboux transformations are used to get some
explicit solutions of the KdVES, which include soliton, rational,
positon, and negaton solutions.}


\Classification{35Q51; 35Q53; 37K10}

\section{Introduction}
The nonlinear evolution equation with sources has many
applications in physics, such as hydrodynamics, plasma physics,
solid state physics (see, e.g.,
\cite{Mel88,Shch96,UK01,Zeng94,Lin01,ZML00,Shao03,Ma2005,Hu96,Zhang03,Ge-Hu06}).
Some $(1+1)$-dimensional systems of this kind can be written in
bi-Hamiltonian form by changing the role of $x$ and $t$
\cite{Zeng99,ZLW98}. In recent years, certain equations with
self-consistent sources have been studied by the inverse
scattering method \cite{Mel88,UK01,Lin01,ZML00}, Darboux
transformation \cite{Shao03,Ma2005}, and Hirota method
\cite{Hu96,Zhang03,Ge-Hu06}. Several types of solutions have been
obtained.

In early works on soliton equations with self-consistent sources
(SESCSs), since the Lax pairs of these systems were not obtained
explicitly, the use of the inverse scattering method was
complicated and needed special skills \cite{Mel88}. Because the
restricted f\/lows of soliton equations are just the stationary
problems of SESCSs and the Lax pairs of restricted f\/lows can
always be deduced from the adjoint representation of soliton
equations \cite{Zeng93}, a natural and simple way to deduce the
auxiliary linear problems of SESCSs was discovered \cite{Zeng94},
and then a systematical approach to solve SESCSs by inverse
scattering method was developed \cite{Lin01,ZML00}.

The Hirota method has been used to construct several types of
solutions for some SESCSs \cite{Hu96,Zhang03}. Recently, X.B.~Hu
and his colleagues developed a new procedure to construct some
systems with self-consistent sources (see, e.g.,~\cite{Ge-Hu06}).
Especially, some discrete systems with sources were studied there.

Darboux transformation is a powerful tool to construct some
solutions of the dif\/ferential equations \cite{Matveev-Salle}.
Some SESCSs have been studied by Darboux transformation (see
\cite{Shao03,Ma2005} and the references therein). In
\cite{Shao03}, the KdV equation with self-consistent sources
(KdVES) is studied by the generalized {\it binary} Darboux
transformation, whose reduction to the form of Darboux
transformation for the original KdV equation ({\it without
source}) is not explicit. Some soliton, positon, and negaton
solutions of the KdVES are obtained in \cite{Shao03}. W.X.~Ma
introduced complexiton solution to some soliton equations and
showed that the complexiton solution of the KdVES can be obtained
by Darboux transformation \cite{Ma2005}. Notice the close relation
between the restricted f\/lows and the SESCSs, and stimulated by
the study on the B\"acklund transformations for the restricted
f\/lows, which were studied on the basis of Darboux--Crum
transformation by V.B.~Kuznetsov
et~al.~\cite{Kuznetsov98,Kuznetsov99,Kuznetsov98new}, two types of
Darboux transformations for the KdVES are constructed in this
paper. The f\/irst type of Darboux transformation is not a {\it
binary} Darboux transformation, but it also enables us to obtain
the soliton and rational solutions of the KdVES. The second type
of Darboux transformation enables us to obtain the positon and
negaton solutions of the KdVES. Some special cases of the
solutions of the KdVES obtained in this paper reduce to some known
solutions in other papers \cite{Lin01,Shao03}.

The paper is organized as follows. In Section 2, the restricted
f\/lows of the KdV hierarchy and the KdV hierarchy with
self-consistent sources are constructed. In Section 3, the Lax
pair for the restricted f\/lows and the auxiliary linear problems
for the KdVES are deduced. In Section~4, two types of Darboux
transformations for the KdVES are constructed, then some solutions
of the KdVES are obtained. Finally, we summarize the main results
of this paper in Section 5.

\section[The restricted flows and the KdVES]{The restricted f\/lows and the KdVES}
Consider the Schr\"{o}dinger equation
\begin{gather}
\label{eqn:Schro}
    \phi_{xx}+(\lambda+u) \phi =0,
\end{gather}
where $\phi$ and $u$ are functions of $x$ and $t$, $\lambda$ is a
spectral parameter. Equation (\ref{eqn:Schro}) can be written in
the matrix form
\begin{gather}
\label{eqn:kdv-spe}
    \left( \begin{array}{c}
    \phi \\ \phi_x \end{array} \right)_x=
    U \left( \begin{array}{c}
    \phi \\ \phi_x \end{array} \right),
    \qquad
    U = \left( \begin{array}{cc}
    0 & 1 \\ -\lambda-u & 0 \end{array} \right).
\end{gather}
The adjoint representation of (\ref{eqn:kdv-spe}) reads
\cite{newell}
\begin{gather}
\label{eqn:p1} V_x=[U, V]\equiv UV-VU.
\end{gather}
Set
\begin{gather*}
%\label{eqn:p2}
    V=\sum_{i=0}^{\infty} \left( \begin{array}{cc}
    a_i & b_i \\ c_i & -a_i \end{array} \right)\lambda^{-i}.
\end{gather*}
Equation (\ref{eqn:p1}) yields
\begin{gather*}
a_0=b_0=0,\quad c_0=-4,\qquad a_1=0,
\qquad b_1=4,\qquad c_1=-2u,\\
a_2=u_x,\qquad b_2=-2u,\qquad c_2=\frac 12(u_{xx}+u^2),\qquad
\dots,
\end{gather*}
and in general for $k=1,2,\dots,$
\begin{gather}
\label{eqn:p3}
 a_k=-\frac{1}{2}b_{k,x},\qquad
    b_{k+1}=Lb_k=-\frac{1}{2}L^{k-1}u,
    \qquad c_k=-\frac{1}{2}b_{k,xx}-b_{k+1}-b_k u,
\end{gather}
where
\[ L=-\frac{1}{4}D^2-u+\frac{1}{2}D^{-1}u_x,
    \qquad D=\frac{\partial}{\partial x},
    \qquad D D^{-1}=D^{-1}D =1. \]

Set
\begin{gather*}
%\label{eqn:p4}
    V^{(n)}=\sum_{i=0}^{n+1} \left( \begin{array}{cc}
    a_i & b_i \\ c_i & -a_i \end{array} \right)\lambda^{n+1-i}
      + \left( \begin{array}{cc}
    0 & 0 \\ b_{n+2} &0 \end{array} \right),
\end{gather*}
and take
\begin{gather}
\label{eqn:p5}
    \left( \begin{array}{c}
    \phi \\ \phi_x \end{array} \right)_{t_n}=
    V^{(n)}(u, \lambda) \left( \begin{array}{c}
    \phi \\ \phi_x \end{array} \right).
\end{gather}
Then the compatibility condition of equations (\ref{eqn:kdv-spe})
and (\ref{eqn:p5}) gives rise to the KdV hierarchy
\begin{gather*}
%\label{eqn:p6}
    u_{t_n}=D\frac{\delta H_n}{\delta u}\equiv
    -2b_{n+2,x}, \qquad n=0,1,\ldots,
\end{gather*}
where $H_n=\frac {4b_{n+3}}{2n+3}$. We have
\begin{gather}
\label{eqn:p7} \frac{\delta \lambda}{\delta u}=\phi^2,\quad\quad
L\phi^2=\lambda\phi^2.
\end{gather}

The high-order restricted f\/lows of the KdV hierarchy consist of
the equations obtained from the spectral problem (\ref{eqn:Schro})
for $N$ distinct $\lambda_j$ and the restriction of the
variational derivatives for conserved quantities $H_n$ and
$\lambda_j$~\cite{Zeng94}
\begin{gather}
\label{eqn:p8}
    \frac{\delta H_n}{\delta u} -
    2\sum_{j=1}^N
    \frac{\delta \lambda_j}{\delta u} \equiv
     -2b_{n+2}-2\sum_{j=1}^N \phi_j^2 =0,
\\
    \phi_{j,xx}+(\lambda_j+u)\phi_j=0,\qquad j=1,\ldots,N,\nonumber
\end{gather}
where $n=0,1,\ldots$.

The KdV hierarchy with self-consistent sources is given
by\cite{Zeng94}
%\begin{subequations}
\begin{gather}
%\label{eqn:p10a}
    u_{t_n}=D\left[ \frac{\delta H_n}{\delta u} -
    2\sum_{j=1}^N
    \frac{\delta \lambda_j}{\delta u} \right]\equiv
    D\left[ -2b_{n+2}-2\sum_{j=1}^N
    \phi_j^2 \right],\label{eqn:p10}
\\
%\label{eqn:p10b}
    \phi_{j,xx}+(\lambda_j+u)\phi_j=0,\qquad j=1,\ldots,N,\nonumber
\end{gather}
where $\lambda_j$ are distinct.

In this paper, we concentrate on the case $n=1$ and denote
$t\equiv t_1$, the equation (\ref{eqn:p10}) gives the KdV equation
with self-consistent sources (KdVES)
\begin{subequations}
\label{eqKdVES}
\begin{gather}
\label{eqKdVESa}
    u_{t}=-(6uu_x+u_{xxx})-
    2\frac{\partial}{\partial x}\sum_{j=1}^N \phi_j^2,
\\
\label{eqKdVESb}
    \phi_{j,xx}+(\lambda_j+u)\phi_j=0,\qquad j=1,\ldots,N.
\end{gather}
\end{subequations}

\section[The Lax pair for the restricted flows and that for the KdVES]{The Lax
pair for the restricted f\/lows and that for the KdVES}

Set $\Phi=(\phi_1,\dots,\phi_N)^T$. According to equations
(\ref{eqn:p3}), (\ref{eqn:p7}) and (\ref{eqn:p8}), we denote
\begin{gather*}
 \widetilde
a_i=a_i,\qquad \widetilde b_i=b_i,\qquad \widetilde c_i=c_i,\qquad
i=0,1,\dots,n+1, \\
\widetilde b_{n+2+i}=-\langle \Lambda^i\Phi,
\Phi\rangle,\qquad i=0,1,2,\dots,\\
\widetilde a_{n+2+i}=-\frac
12\widetilde b_{n+2+i,x}=\langle \Lambda^i\Phi, \Phi_x\rangle ,\\
\widetilde c_{n+2+i}=-\frac 12\widetilde b_{n+2+i,xx}-\widetilde
b_{n+3+i}-\widetilde b_{n+2+i}u=\langle \Lambda^i\Phi_x,
\Phi_x\rangle.
\end{gather*}
Then
\begin{gather*}
N^{(n)}\equiv \left( \begin{array}{cc}
    A^{(n)} & B^{(n)} \\C^{(n)} & D^{(n)}\end{array} \right)
 =\lambda^{n+1} \sum_{k=0}^{\infty} \left( \begin{array}{cc}
    \widetilde a_k &\widetilde  b_k \\\widetilde c_k & -\widetilde a_k \end{array} \right)
    \lambda^{-k} \\
\phantom{N^{(n)}}{} = \sum_{k=0}^{n+1} \left( \begin{array}{cc}
    a_k & b_k \\ c_k & -a_k \end{array} \right)
    \lambda^{n+1-k}
     + \sum_{j=1}^N \frac{1}{\lambda-\lambda_j}
    \left( \begin{array}{cc}
    \phi_j\phi_{j,x} & -\phi_j^2 \\
    \phi_{j,x}^2 & -\phi_j\phi_{j,x}
    \end{array} \right),
\end{gather*}
also satisf\/ies the adjoint representation (\ref{eqn:p1}), i.e.\
\begin{gather}
\label{eqn:p9} N^{(n)}_x=[U, N^{(n)}].
\end{gather}
In fact equation (\ref{eqn:p9}) gives rise to the Lax
representation of the restricted f\/low (\ref{eqn:p8}).

Since the high-order restricted f\/lows (\ref{eqn:p8}) are just
the stationary equations of the KdV hierarchy with self-consistent
sources (\ref{eqn:p10}), it is obvious that the zero-curvature
representation for the KdV hierarchy with self-consistent sources
(\ref{eqn:p10}) is given by
\begin{gather*}
%\label{ss1}
    U_{t_n}-N^{(n)}_x+ [U, N^{(n)}]=0,
\end{gather*}
with the auxiliary linear problems
\begin{gather*}
%\label{eqn:KdVH-Source-Zero}
    {\left( \begin{array}{c}
    \psi \\ \psi_x \end{array} \right)}_x
    = U {\left( \begin{array}{c}
    \psi \\ \psi_x \end{array} \right)},\qquad
{\left( \begin{array}{c}
    \psi \\ \psi_x \end{array} \right)}_{t_n}
    = N^{(n)} {\left( \begin{array}{c}
    \psi \\ \psi_x \end{array} \right)},
\end{gather*}
or equivalently
\begin{subequations}
\label{KdVHWS-Lax}
\begin{gather}
\label{KdVHWS-Laxa}
    \psi_{xx}+(\lambda+u) \psi =0,
\\
\label{KdVHWS-Laxb}
   \psi_{t_n}=A^{(n)}\psi+B^{(n)}\psi_x
     = \sum\limits_{l=0}^{n+1}(a_l\psi+b_l\psi_x)\lambda^{n+1-l}+
    \sum\limits_{j=1}^N \frac{1}{\lambda-\lambda_j}
    \phi_j(\phi_{j,x}\psi-\phi_j\psi_x).
\end{gather}
\end{subequations}
In particular, the system (\ref{KdVHWS-Lax}) for $n=1$ gives the
auxiliary linear problem for the KdVES~(\ref{eqKdVES})
\begin{subequations}
\label{KdVES-Lax}
\begin{gather}
\label{KdVES-Lax-a} \psi_{xx}+(\lambda+u) \psi =0,
\\
\label{KdVES-Lax-b} \psi_{t}= u_x \psi +(4\lambda-2u)\psi_x+
\sum_{j=1}^N \frac{1}{\lambda-\lambda_j}
\phi_j(\phi_{j,x}\psi-\phi_j\psi_x).
\end{gather}
\end{subequations}
The compatibility condition of (\ref{KdVES-Lax-a}) and
(\ref{KdVES-Lax-b}) gives the KdVES (\ref{eqKdVES}) under the
assump\-tion~(\ref{eqKdVESb}).

\section{The Darboux transformation for the KdVES}

The B\"acklund transformation for the restricted f\/lows of KdV
hierarchy (\ref{eqn:p8}) has been studied by Kuznetsov et
al.~\cite{Kuznetsov98,Kuznetsov99,Kuznetsov98new}. These
transformations can be extended to construct the Darboux
transformation for the KdVES. Two types of Darboux transformations
and some solutions for the KdVES will be constructed in this
section.

In the following, we use $W(g_1,\dots,g_m)$ to denote the
Wronskian determinant for functions $g_1(x), g_1(x), \dots,
g_m(x)$, i.e.,
    \[ W(g_1,g_2,\dots,g_m)=\left|\begin{array}{cccc}
    g_1 & g_2 & \cdots & g_m \\
    \partial_x g_{1} & \partial_x g_{2} & \cdots & \partial_x g_{m} \\
    \cdots & \cdots & \cdots & \cdots \\
    \partial_x^{m-1} g_1 & \partial_x^{m-1} g_2 & \cdots & \partial_x^{m-1} g_m
    \end{array}\right|. \]

\subsection{First type of Darboux transformation for the KdVES}

\begin{proposition}
\label{Prop-1} Suppose $u, \phi_1, \ldots, \phi_N, $ is a solution
of the KdVES \eqref{eqKdVES}, $\psi, u, \phi_1, \ldots, \phi_N, $
sa\-tis\-fy the linear problem \eqref{KdVES-Lax}. If the functions
$f(x,t,\lambda_{N+1})$ and $g(x,t,\lambda_{N+1})$ are two
solutions of~\eqref{KdVES-Lax} with $\lambda=\lambda_{N+1}$ (where
$\lambda_{N+1}\neq\lambda_j$ for $j=1,\ldots,N$), and $W(f,g)\neq
0$, then the following functions satisfy the linear problem
\eqref{KdVES-Lax} (with $N$ replaced by $N+1$)
\begin{gather}
\label{Darboux-1}
    \widetilde \psi = \frac{W(S,\psi)}{S},
    \qquad \widetilde u = u+2 \partial_x^2 \ln S,
\\
\widetilde \phi_j = \frac{1}{\sqrt{\lambda_j-\lambda_{N+1}}}
\frac{W(S,\phi_j)}{S},
            \qquad j=1,\ldots,N, \qquad
 \widetilde \phi_{N+1} = \sqrt{\frac{C_t}{W(f,g)}} \frac{W(S,f)}{S},\nonumber
\end{gather}
where $S=C(t) f(x,t,\lambda_{N+1}) + g(x,t,\lambda_{N+1})$, $C(t)$
is a differentiable function of $t$, and $C_t=\frac{d C}{dt}$.
\end{proposition}

\begin{proof}
It can be proved by direct computation.
\end{proof}

It is easy to check that $\widetilde \phi_j$ ($j=1,\ldots,N+1$) in
(\ref{Darboux-1}) satisf\/ies (\ref{eqKdVESb}) under the
assumption in Proposition~\ref{Prop-1}. That means, we can use the
Darboux transformation (\ref{Darboux-1}) (with $C(t)$ being
variant in $t$) to obtain a solution of the KdVES (\ref{eqKdVES})
with $N$ replaced by $N+1$. If we f\/ix the function $C(t)$ in
(\ref{Darboux-1}) to be a {\it constant}, then the Darboux
transformation (\ref{Darboux-1}) for the case $N=0$ reduces to the
Darboux transformation for the original KdV equation ({\it without
source}).

\subsubsection{Soliton solution}

It is easy to see that the KdVES (\ref{eqKdVES}) with $N=1$ and
$\lambda_1=0$ has the following solution
\begin{gather*}
    u=0,\qquad
    \phi_1=\eta(t).
\end{gather*}
With the above $u$ and $\phi_1$, we take two solutions of
(\ref{KdVES-Lax}) for $\lambda=-k^2$ (where $k>0$) as
    \[ f = \exp(kx-a(t)), \qquad g = \exp(-kx+a(t)),\]
where $a(t)$ is a dif\/ferentiable function of $t$ and
\begin{gather}
\label{soliton-dadt}
    \frac{da}{dt}=4k^3-\frac{\eta(t)^2}{k}.
\end{gather}
Then use the Darboux transformation (\ref{Darboux-1}) with
$C(t)=\exp(-2z(t))$, where $z(t)$ is a dif\/feren\-tiable function
of $t$, we get a solution of the KdVES (\ref{eqKdVES}) with $N=2$
\begin{gather}
\label{soliton}
    \widetilde u=2 k^2 \,\mbox{sech}^2 (kx-a(t)-z(t)),
    \qquad
    \widetilde \phi_1=-\eta(t) \tanh(kx-a(t)-z(t)),
\\
    \widetilde \phi_2= \sqrt{k \frac{d z}{d t}} \,\mbox{sech}\,(kx-a(t)-z(t)),\nonumber
\end{gather}
where $a(t)$ is given by (\ref{soliton-dadt}). The velocity of
propagation of this soliton solution can be modif\/ied by the
choice of the function $z(t)$ \cite{Mel88,Lin01}. This phenomena
is dif\/ferent with the case of the original KdV equation where
the velocity is proportional to the amplitude of soliton. One
should notice that $|\widetilde \phi_1| \rightarrow |\eta(t)|$
when $|x|\rightarrow \pm \infty$, this kind of source is studied
less in the references. In addition, if we set $\eta(t)\equiv 0$,
the solution (\ref{soliton}) reduces to the soliton solution
obtained in some other papers \cite{Mel88,Lin01}.

\subsubsection{Rational solution}

The KdVES (\ref{eqKdVES}) with $N=0$ has a trivial solution $u=0$.
Take two solutions of (\ref{KdVES-Lax}) with $u=0$ and $\lambda=0$
as follows
    \[ f = 1, \qquad g = x,\]
then use the Darboux transformation (\ref{Darboux-1}), we get a
solution of the KdVES (\ref{eqKdVES}) with $N=1$
\begin{gather*}
%\label{rational}
    \widetilde u=\frac {-2}{(x+C(t))^2},
    \qquad
    \widetilde \phi_1=\frac {-\sqrt{C_t}}{(x+C(t))}.
\end{gather*}
It is a rational solution, and the poles of the solution are
variant with respect to the choice of function $C(t)$.

\subsection{Second type of Darboux transformation for the KdVES}

In the following, we will construct another type of Darboux
transformation for the KdVES (\ref{eqKdVES}), which enables us to
obtain positon and negaton solutions of the KdVES (\ref{eqKdVES}).

\begin{proposition}
\label{Prop-2} Suppose $u, \phi_1, \dots , \phi_N, $ is a solution
of the KdVES \eqref{eqKdVES}, $\psi, u, \phi_1, \dots , \phi_N, $
sa\-tis\-fy the linear problem \eqref{KdVES-Lax}. If the functions
$f(x,t,\lambda_{N+1})$ and $g(x,t,\lambda_{N+1})$ are two
solutions of~\eqref{KdVES-Lax} with $\lambda=\lambda_{N+1}$ (where
$\lambda_{N+1}\neq\lambda_j$ for $j=1,\dots ,N$), and $W(f,g)\neq
0$), then the following functions satisfy the linear problem
\eqref{KdVES-Lax} (with $N$ replaced by $N+1$)
\begin{gather}
\label{Darboux-2}
    \widetilde \Psi = \frac{W(g,T,\Psi)}{W(g,T)},
    \qquad \widetilde u = u+2 \partial_x^2 \ln W(g,T),
\\ \widetilde \phi_j = \frac{1}{\lambda_j-\lambda_{N+1}}
        \frac{W(g,T,\phi_j)}{W(g,T)},
        \qquad j=1,\dots ,N, \qquad
 \widetilde \phi_{N+1} = \sqrt{\frac{C_t}{W(f,g)}}
        \frac{W(g,T,f)}{W(g,T)}, \nonumber
\end{gather}
where $T=C(t)f(x,t,\lambda_{N+1})+\partial_{N+1}
g(x,t,\lambda_{N+1})$ and $C(t)$ is a differentiable function of
$t$.
\end{proposition}

\begin{proof}
It can be proved by direct computation.
\end{proof}

It is easy to check that $\widetilde \phi_j$ ($j=1,\dots ,N+1$) in
(\ref{Darboux-2}) satisf\/ies (\ref{eqKdVESb}) under the
assumption in Proposition \ref{Prop-2}. That means, we can use the
Darboux transformation (\ref{Darboux-1}) (with $C(t)$ being
variant in $t$) to obtain a solution of the KdVES (\ref{eqKdVES})
with $N$ replaced by $N+1$.

\subsubsection{Positon solution}

The KdVES (\ref{eqKdVES}) with $N=1$ and $\lambda_1=0$ has a
trivial solution
    \[ u=0, \qquad \phi_1=\sqrt{\frac{d\eta(t)}{dt}},
    \]
where $\eta(t)$ is a dif\/ferentiable function of $t$. With the
above $u$ and $\phi_1$, we take two solutions of~(\ref{KdVES-Lax})
for $\lambda=k^2$ (where $k>0$) as
    \[ f = \cos\Theta, \qquad g = \sin\Theta,
    \qquad \Theta=kx+4k^3t-\frac{\eta(t)}{k}+b(k), \]
where $b(k)$ is a dif\/ferentiable function of $k$. By using the
Darboux transformation (\ref{Darboux-2}), we get a solution of
KdVES (\ref{eqKdVES}) with $N=2$
\begin{gather}
\label{Positon}
    \widetilde u=\frac
  {32 k^2(2k^2 \gamma \cos\Theta-\sin\Theta)\sin\Theta
  }
  {
  (4k^2 \gamma -\sin(2\Theta))^2
  },
\\
 \widetilde \phi_1=   \frac
    {
    -\sqrt{\eta_t} (4 k^2 \gamma +\sin(2\Theta))
    }
    {
    4k^2 \gamma -\sin(2\Theta)
    }, \qquad
    \widetilde \phi_2=  \frac
    {
    4 k \sqrt{k C_t} \sin\Theta
    }
    {
    4k^2 \gamma -\sin(2\Theta)
    }, \nonumber
\end{gather}
where
    \[ \gamma=C(t)+\frac{1}{2k} \partial_k \Theta. \]
This is a positon solution (see \cite{Shao03} and the references
therein). If we set $\phi_1=0$, then the
solu\-tion~(\ref{Positon}) reduces to the one given in
\cite{Shao03}.

\subsubsection{Negaton solution}

The KdVES (\ref{eqKdVES}) with $N=1$ and $\lambda_1=0$ has a
trivial solution
    \[ u=0, \qquad \phi_1=\sqrt{\frac{d\eta(t)}{dt}},\]
where $\eta(t)$ is a dif\/ferentiable function of $t$. With the
above $u$ and $\phi_1$, we take two solutions of~(\ref{KdVES-Lax})
for $\lambda=-k^2$ (where $k>0$) as
    \[ f = \cosh\Theta, \qquad g = \sinh\Theta,
    \qquad \Theta=kx-4k^3t+\frac{\eta(t)}{k}+b(k), \]
where $b(k)$ is a dif\/ferentiable function of $k$. By using the
Darboux transformation (\ref{Darboux-2}), we get a solution of
KdVES (\ref{eqKdVES}) with $N=2$
\begin{gather}
\label{Negaton}
    \widetilde u=\frac
  {8 k^2 (2k^2\gamma\cosh\Theta + \sinh\Theta) \sinh\Theta
  }
  {
  (2k^2\gamma + \sinh\Theta \cosh\Theta)^2
  },
\\
    \widetilde \phi_1=  \frac
    {
    \sqrt{\eta_t}(-2k^2\gamma +\sinh\Theta \cosh\Theta)
    }
    {
    2k^2\gamma + \sinh\Theta \cosh\Theta
    },    \qquad
    \widetilde \phi_2=  \frac
    {
    2 k \sqrt{k C_t} \sinh\Theta
    }
    {
    2k^2\gamma + \sinh\Theta \cosh\Theta
    },\nonumber
\end{gather}
where
    \[ \gamma=C(t)-\frac{1}{2k} \partial_k \Theta. \]
This is a negaton solution (see \cite{Shao03} and the references
therein). If we set $\phi_1=0$, then the solution~(\ref{Negaton})
reduces to the one given in \cite{Shao03} (notice that there is a
typo in the expression of negaton solution (4.2b) in
\cite{Shao03}).

\section{Conclusion}

The KdV equation is used as an example to illustrate the relation
between the restricted f\/lows and the soliton equation with
self-consistent sources (SESCSs). Since the restricted f\/lows is
just the stationary problem of the SESCSs and the Lax pair of the
restricted f\/lows can always be deduced from the adjoint
representation of soliton equations, the auxiliary linear problem
for the SESCSs can be easily obtained. Stimulated by the study on
the B\"acklund transformation for the restricted f\/lows (by
Kuznetsov et al.) \cite{Kuznetsov98,Kuznetsov99,Kuznetsov98new},
two types of Darboux transformations  for the KdVES are
constructed in this paper. The f\/irst type of Darboux
transformation is not a {\it binary} one, whose reduction relation
to the form of Darboux transformation for the original KdV
equation is shown. By the two types of Darboux transformation,
some explicit solutions for the KdVES are obtained, which include
soliton, rational, positon, and negation solutions. It is possible
to construct complexiton solution of the KdVES by the Darboux
transformation in this paper, we will study it in detail in the
future work.

\bigskip

{\it This paper is submitted to the memorial volume for Vadim B.
Kuznetsov. Dr. Kuznetsov was an expert in integrable systems. One
of the authors (R.L. Lin) visited him at Leeds in 2002 at his
invitation. R.L. Lin was deeply impressed that he was so kindly to
help the younger researchers. As we know, he kept close contact
with some Chinese researchers and he visited another author (Y.B.
Zeng) at Tsinghua University in 2002. Also, he applied some
financial aid to help several Chinese researchers to work in
Leeds. Vadim left, but his smile is left in our memory. }

\subsection*{Acknowledgements}

The authors are grateful to the referees for the valuable
comments. This work is supported by the Chinese Basic Research
Project ``Nonlinear Science". R.L.~Lin is supported in part by
``Scientif\/ic Foundation for Returned Overseas Chinese Scholars,
Ministry of Education".

\LastPageEnding

\end{document}